\documentstyle[prb,aps,eqsecnum,twocolumn,epsf]{revtex}

\newcommand{\beq}{\begin{equation}}
\newcommand{\eeq}{\end{equation}}
\newcommand{\bea}{\begin{eqnarray}}
\newcommand{\eea}{\end{eqnarray}}

\newcommand{\slwc}{checkerboard lattice}

\newcommand{\etal}{{\em et al.}}

\newcommand{\bu}{b_\uparrow}
\newcommand{\bd}{b_\downarrow}
\newcommand{\da}{^\dagger}

\def\tit#1#2#3#4#5{{#1} {\bf #2}, #3 (#4)}

\def\prl{Phys.\ Rev.\ Lett.\ }
\def\pr{Phys.\ Rev.\ }
\def\prb{Phys.\ Rev.\ B\ }

\def\jap{J.\ Appl.\ Phys.\ }
\def\zpb{Z.\ Phys.\ B\ }
\def\jpsj{J.\ Phys.\ Soc.\ Jpn.\ }

\def\ijmpb{Int.\ J.\ Mod.\ Phys.\ B\ }

\def\unue#1{\noindent{\bf #1}}

\begin{document}
\draft

\twocolumn[\hsize\textwidth\columnwidth\hsize\csname @twocolumnfalse\endcsname

\title{
Planar pyrochlore, quantum ice and sliding ice
} 
\author{R. Moessner,$^1$ Oleg Tchernyshyov,$^2$ and S. L. Sondhi$^1$}
\address{$^1$Department of Physics, Princeton University,
Princeton, NJ 08544\\
$^2$School of Natural Sciences,
Institute for Advanced Study, Princeton, NJ 08540} 
\date{June 11, 2001}

\maketitle

\begin{abstract}
We study quantum
antiferromagnetism on the highly frustrated \slwc, also known as 
the square lattice with crossings. The quantum Heisenberg antiferromagnet 
on this lattice is of interest as a two-dimensional analog of the pyrochlore 
lattice magnet. By combining several approaches we conclude that this system
is most likely ordered for all values of spin, $S$, with a N\'eel
state for large $S$\ giving way to a two-fold degenerate valence-bond 
solid for smaller $S$. We show next that the Ising antiferromagnet with a 
weak four-spin exchange, equivalent to square ice with the leading
quantum dynamics, exhibits long range ``anti-ferroelectric'' order. As a
byproduct of this analysis we obtain, in the system of weakly coupled
ice planes, a sliding phase 
with XY symmetry.

\end{abstract}
]

\mbox{}
\vspace{0.cm}

\unue{Introduction:} On the heels of recent progress in understanding
highly frustrated classical magnets, coupled with a substantial
experimental effort,\cite{hfmrev} renewed attention is now focused on
the behaviour of their quantum counterparts.  In particular, the
Heisenberg pyrochlore antiferromagnet is being studied with view to
the question of whether frustration-enhanced quantum fluctuations
might lead to unconventional ordering -- or complete absence thereof
-- especially for small values of the quantum spin,
$S$.\cite{harber,isodapyro,canalspyro,ga-h,kogaseries,tsunetsugu,elhajal}
This model classically is special in that frustration prevents any
sort of ordering or dynamical phase transition down to the lowest
temperatures,\cite{villain,pyrodisordered,pyroshlo} for which reason
it is termed a cooperative paramagnet or classical spin liquid.

The challenge of this problem arises from the small energy scale
generated by the frustration: in a semiclassical picture, any linear
combination of the classically degenerate ground states -- the
collection of which is of extensive dimensionality\cite{pyroshlo}
 -- may be selected 
as the quantum ground state.
For the highly frustrated two-dimensional magnet on the
related kagome lattice, exact diagonalisations of small
clusters\cite{kagdiag} have provided crucial benchmarks. This system
has turned out to be particularly well suited to this approach as it
appears to have a very short correlation length, although one does
find a large number of low-lying singlet excitations.

For the pyrochlore magnet, it looks as if such results will elude us
for some time to come. The pyrochlore lattice, being
three-dimensional, displays a more inclement scaling of the Hilbert
space dimension with linear system size.  Moreover, its unit cell
contains four spins and its structure implies that the
smallest system without spurious boundary condition effects contains
at least 16 sites.

\begin{figure}
\vspace{-1cm}
\epsfxsize=3.1in
\centerline{\epsffile{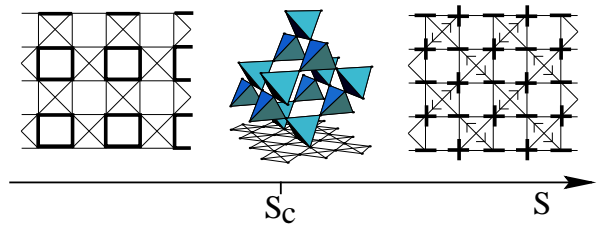}}
\caption{ Phase diagram of the \slwc\ Heisenberg magnet as a function
of spin, $S$.  The plaquette (left -- fat bonds are those with
enhanced probability of singlet formation) and N\'eel
(antiferroelectric) phases. A projection of the 3-d pyrchlore lattice
gives the 2-d \slwc\ (middle). A tetrahedron becomes a square with
crossings upon projection, all bonds of which have equal strength. }
\label{fig:sqcrossofS} 
\end{figure}

To evade this, attention has shifted to a system which avoids some of
these complications, namely the \slwc\ (Fig.~\ref{fig:sqcrossofS}).
It is expected to have similar properties to the pyrochlore as it has
the same local structure -- both can be thought of as networks of
corner-sharing tetrahedra. Further, the size and topology of its
ground state manifold for Heisenberg magnets are identical to the
pyrochlore case.\cite{pyroshlo} However, it has a unit cell of only
two spins, is two dimensional, and has allowed exact diagonalisations
of a good number of finite size systems of up to 36 spins.  Such
diagonalisations have very recently been carried out by Palmer and
Chalker\cite{palcha} and by Fouet \etal \cite{fouetslwc} and other
workers have also recently studied this system by several
techniques.\cite{liebschupp,canalsinv,canalswave,elhajal}

In this paper, we take a somewhat broader view of quantum
antiferromagnetism on the \slwc. For the Heisenberg problem, apart
from the small $S$ cases, which we study by a dimer model analysis and
an Sp$(N)$ mean field theory, we analyze the large $S$ region, both
within Sp$(N)$ -- which is also able to treat the intermediate region
-- and through the semiclassical $1/S$ expansion.  Here we find a
strong ordering tendency -- into a two-fold degenerate valence bond
crystal for small $S$ and a N\'eel state at large $S$
(Fig.~\ref{fig:sqcrossofS}).  Our predictions find support in the
numerical work of Fouet \etal\ as we discuss further below. The nature
of the valence bond order is at odds with other recent
work.\cite{elhajal}
The $S=1/2$\ kagome magnet\cite{kagdiag}  behaves very differently 
from the checkerboard -- 
and therefore probably the pyrochlore -- ones. Details of
the ordering, however, depend on properties of the \slwc\ (most
prominently, the explicit breaking of the symmetry between bonds in a
tetrahedron and the existence of nontrivial closed loops of length
four residing on a bipartite lattice) which it does not share with the
pyrochlore lattice; there we expect the order to be much more
delicate.

In addition we consider the Ising antiferromagnet with a weak four
spin exchange dynamics -- the Ising ground states are isomorphic to
those of square ice and the dynamics represent the shortest ring
exchanges in this manifold arising from quantum tunneling -- hence
``quantum ice''. The resulting state displays long range
``antiferroelectric'' order in ice terminology, or N\'eel order in
Ising terminology (Fig.~\ref{fig:sqcrossofS}). A byproduct of this
analysis is the state of a stack of weakly coupled ice layers, which is
seen to be in a ``sliding'' phase of the kind discussed in recent
work,\cite{floatingphase} albeit one that is protected by much simpler
arrangements.

We turn now to the details of these assertions and sketches of the
underlying analyses.  We treat the Heisenberg problem first.

\unue{Sp($N$) generalization:} The Sp($N$) technology for frustrated
magnets, a reformulation of Schwinger boson mean field
theory\cite{arauer} controlled by the introduction of $1/N$ as a small 
parameter, was introduced and described in detail by Read and
Sachdev.\cite{readsachsp} 
One starts by rewriting the SU(2)$\sim$Sp(1) 
spin operators in terms of bosonic operators,
$\left\{\bu,\bd\right\}$, with the constraint of
$\bu\da\bu+\bd\da\bd=2S$\ on each site, and
${S}^z=(\bu\da\bu-\bd\da\bd)/2$, 
${S}^+=\bu\da\bd$. 
The antiferromagnetic nearest-neighbour Heisenberg Hamiltonian
is rewritten in terms of the bosonic operators (up to a constant):
${H}=J\sum_{\left<ij\right>}{\mathbf{S_i\cdot S_j}}=
(-J/2)\sum_{\left<ij\right>}
(\epsilon^{\sigma\tau}b\da_{i\sigma}b\da_{j\tau})
(\epsilon^{\mu\nu}b_{i\mu}b_{j\nu})$. One generalises this
expression to Sp($N$) by formally introducing $N$\ flavours of
bosons, labelled by capital letters, and replacing the antisymmetric
tensor in ${H}$\ by its Sp($N$) generalisation ${\cal
J}^{\mu\nu}_{AB}=\epsilon^{\mu\nu}\delta_{AB}$.
In the limit
$N\rightarrow\infty$\ at a fixed boson number per flavour per site,
$\kappa$, one obtains to leading order 
in $1/N$ a 
mean-field theory for spin $S=\kappa/2$;
fluctuations about this give rise to a gauge theory.

Self-consistent solutions to the mean field theory are obtained by
minimising with respect to the Hubbard-Stratonovich link fields
$Q_{ij}=\left< {\cal J}^{\mu\nu}_{AB}b_{i\mu}^{A}b_{j\nu}^{B}\right>$,
subject to a constraint on the boson number $\kappa$.  In the process, one
obtains a dispersion relation for the energies $\omega$ of bosonic
spin 1/2 quasiparticles. Within the mean-field theory, there is
generically a disordered phase at small $\kappa$; when one of the
bosonic 
modes goes soft as $\kappa$\ is increased, condensation of these
`spinons' occurs, and long-range spin order ensues. 

Our results at mean-field level are readily summarised. We find zero
expectation value of the diagonal bond variables at all $\kappa$. On
the remaining bonds, long-range N\'eel order 
(Fig.~\ref{fig:sqcrossofS}) develops above $\kappa_c =
0.393$, exactly as for the simple square lattice.
For $\kappa<\kappa_c$, the N\'eel correlations are only
short-ranged.

Whether or not the spinon excitations in the disordered, small
$\kappa$ phase remain deconfined can only be settled by going beyond
mean-field theory. As on the square lattice, one obtains a compact
$U(1)$\ gauge theory at $O(1/N)$, in which instanton tunneling effects
lead to the formation of bond (Peierls) order.\cite{readsachsp} An
analogous calculation is presented by Chung \etal\ for the
Shastry-Sutherland model.\cite{chung} Details of the ordering pattern,
in particular those due to the inequivalence of the plaquettes with
and without crossings, are perhaps most easily studied through
a quantum dimer model (QDM).

\begin{figure}
\epsfxsize=3.3in
\centerline{\epsffile{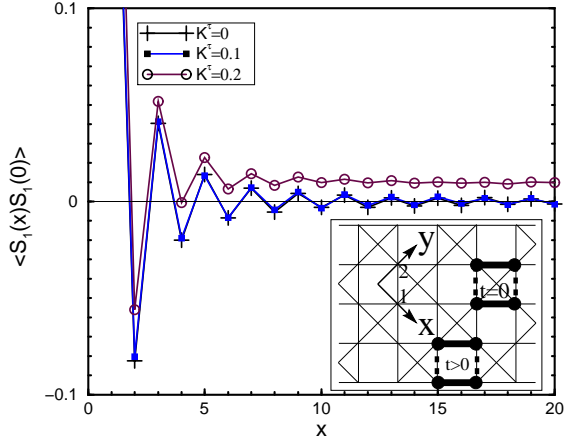}}
\caption{Spin 
correlations on sublattice 1 in the x-direction from Monte Carlo
on
$48\times48\times16$ plaquettes. The classical curve, $K^\tau=0$, and
$K^\tau=0.1$\ coincide. N\'eel long-range order, albeit weak, obtains
for $K^\tau=0.2$. Inset: Numbers label sublattices, arrow translation
vectors. In the QDM, one type of resonance move has zero kinetic energy,
$t=0$. }
\label{fig:SxofT} 
\end{figure}

\unue{Quantum dimer model:} 
This approach starts from the assumption that the
magnet is in a regime where the N\'eel state is destabilised and the
effective degrees of freedom are singlet bonds between 
neighbouring spins, also called valence bonds, which are represented
by dimers.  As each spin participates in exactly one singlet bond, the
Hilbert space consists of all hardcore dimer coverings. An
effective dimer Hamiltonian is obtained by means of an overlap
expansion, described in Ref.~\onlinecite{Rokhsar88}. It is formally
perturbative in a small parameter arising from the non-orthogonality
of the spin wavefunctions describing the different dimer coverings. To
zeroth order, all of the exponentially numerous dimer coverings are
degenerate, but at the next order, a resonance term is generated.
One obtains a
resonance loop of length 4
corresponding to flipping a pair of dimers by $90^\circ$
(Fig.~\ref{fig:SxofT}):
\setlength{\unitlength}{3947sp}%
\begingroup\makeatletter\ifx\SetFigFont\undefined%
\gdef\SetFigFont#1#2#3#4#5{%
  \reset@font\fontsize{#1}{#2pt}%
  \fontfamily{#3}\fontseries{#4}\fontshape{#5}%
  \selectfont}%
\fi\endgroup%
\begin{picture}(154,155)(397,321)
\thicklines
\put(527,452){\circle{18}}
\put(528,344){\line( 0, 1){105}}
\put(527,345){\circle{18}}
\put(421,345){\circle{18}}
\put(421,344){\line( 0, 1){105}}
\put(421,452){\circle{18}}
\end{picture}
$\leftrightarrow$
\setlength{\unitlength}{3947sp}%
\begingroup\makeatletter\ifx\SetFigFont\undefined%
\gdef\SetFigFont#1#2#3#4#5{%
  \reset@font\fontsize{#1}{#2pt}%
  \fontfamily{#3}\fontseries{#4}\fontshape{#5}%
  \selectfont}%
\fi\endgroup%
\begin{picture}(155,154)(533,319)
\thicklines
\put(664,343){\circle{18}}
\put(556,342){\line( 1, 0){105}}
\put(557,343){\circle{18}}
\put(557,449){\circle{18}}
\put(556,449){\line( 1, 0){105}}
\put(664,449){\circle{18}}
\end{picture}.
The
question is what kind of quantum dimer state is selected by this
resonance move.

The most important difference between the well-studied square lattice
QDM and that on the \slwc\ is the following.\cite{fn-complete} As the
overlap expansion is essentially organised by length, $L$, of the
resonance loops, the leading order terms arise from the shortest
possible resonance loops, namely those of length four. Carrying out
the leading order overlap expansion, we find that all the loops of
length four on plaquettes {\em with} crossings (the projected
tetrahedra) have zero kinetic energy, $t$, in the dimer model
and hence play no role. By contrast, the resonance loops on plaquettes
{\em without} crossings, have nonzero kinetic energy 
(Fig.~\ref{fig:SxofT}).
Note that such loops of length four are absent from the pyrochlore
lattice, where the shortest loop contacting more than one tetrahedron
has length six.

Extending the results from the square lattice QDM,\cite{Rokhsar88} one
therefore expects the resulting state to be a valence-bond crystal;
the staggering of $t$\ strongly favours the plaquette crystal, with
the dimers resonating on one of two sublattices of plaquettes
without crossings (Fig.~\ref{fig:sqcrossofS}). Note that the
degeneracy of this state is two rather than four, as would be the case
for the plaquette state on the square lattice, because of the explicit
symmetry breaking introduced by the presence of the plaquettes with
crossing interactions.

\unue{Semiclassics:} We now consider the
case of large spin $S$\ and large $\kappa$\ for SU(2) and Sp($N$),
respectively. In either case, one compares the zero-point energy due
of the excitations (spin waves and spinons, respectively) of
different, classically degenerate ground states. This, in principle,
requires evaluating that energy for the entire ground state manifold,
which is of extensive dimensionality.

For Sp($N$), we have compared the zero-point energies of all
four-sublattice states, as well as of eight-sublattice coplanar
states, and found that the N\'eel state selected at $\kappa_c$\ is
also favoured in this limit, which suggests its stability for all
$\kappa>\kappa_c$.

For large $S$, we have computed the zero-point energy of all
four-sublattice states. We find good qualitative agreement with
Henley's suggestion\cite{hencjp} of an effective energy functional of the
biquadratic type, $-\sum_{\left<ij\right>} ({\bf S}_i\cdot{\bf
S}_j)^2$, in that it correctly reproduces the location of
the maximum and the minima.
There remains a degeneracy between some inequvivalent
collinear states.\cite{hencjp} One of the remaining degenerate states
is indeed the N\'eel state, but other states, disfavoured at
$O(1/\kappa)$, still have exactly the same energy of zero-point
fluctuations at $O(1/S)$.

\unue{Implications for the 3-d pyrochlore magnet:} 
The actual ordering behaviour of the pyrochlore is
still far from settled. At this stage\cite{harber,isodapyro,canalspyro,ga-h,kogaseries,tsunetsugu,elhajal,hencjp}
it appears to be somewhat closer to that of the \slwc\ than
either of the two is to
the kagome case, where there has so far been no
strong indication of long-range order of any kind for $S=1/2$, 
and where the large-$S$ state is necessarily non-collinear.  We
emphasize, however, that details of the ordering we find, such as the
pattern of the bond solid or the size of the unit cell, crucially
depend on differences between the two lattices -- such as spatial 
dimensionality, the presence of inequivalent links, and short closed
loops linking different tetrahedra. 
Moreover, we remark that 
most approaches employed so far for $d=3$\ 
pyrochlore\cite{harber,isodapyro,canalspyro,kogaseries,tsunetsugu,elhajal} 
explicitly remove the equivalence of all tetrahedra by weakening
the bonds on half of them; from such a starting point, it would 
would seem 
rather difficult to restore this equivalence, which is necessary for
obtaining  the plaquette ordering we 
find.

\unue{Finite-size diagonalisations:} The agreement
of the above calculations on a number of
central points is reassuring. Most importantly, they all
predict ordered states with translational symmetry breaking and a
two-fold degeneracy.  For small $S$, they lead us to expect dimer
order of the plaquette flavour, which gives way to a N\'eel
state at intermediate $S$. How does this compare to
exact diagonalisations of $S=1/2$ Heisenberg magnets?

Palmer and Chalker, who have studied systems containing up to 24
spins, find no clearly identifiable degeneracy. Rather, there appears
to be a large number of low-lying singlet states with a small gap and
a much larger gap to triplet excitations. They rule out N\'eel order but
are inconclusive about translational symmetry breaking. 
Fouet \etal\ agree with their
results, but have in addition studied a system with 36 sites. There,
they find a particularly low ground state energy, suggesting that the
boundary conditions for this system size accomodate well the quantum
ground state. For this system, there does appear a two-fold near
degeneracy of the ground state, with the states being described by the
wavevectors expected for our dimer crystal. 

\unue{Quantum ice and sliding ice:} We turn next to the Ising problem.
The ground states of the Ising
antiferromagnet on the \slwc\ require two up and two down spins
on each tetrahedron (square with crossings). The six such
possible configurations on each tetrahedron can be identified with 
the allowed vertices of the six-vertex model as 
follows.\cite{andersonpyro}
Divide the square lattice consisting of the crossings at the
centres of the tetrahedra into sublattices A and B in the usual
fashion. Orient the links coming out of sublattice A(B) inwards
if the spin sitting on that link is up(down) and outwards if
the spin in down(up) (Fig.~\ref{fig:sqcrossofS}). 
As all vertices are weighted equally, the
ground state manifold has the extensive entropy, $(3/4)\ln(4/3)$\ per 
spin, of square ice.\cite{liebice} We will shortly find it useful
that the ice problem also has a height representation in which
an integer valued height living on the dual lattice steps up(down) by 
one on crossing an in(out) arrow clockwise around any 
vertex.\cite{abraham}

As a matter of principle, the degeneracy of the ice manifold will
be lifted by quantum effects which will involve (in six vertex
language) closed loops of arrows that will reverse direction. 
Here we
study the ordering produced by this dynamics. The simplest 
such process for square ice involves a loop around a single plaquette.
Translated into the Ising spin representation on the \slwc\ it gives rise 
to a Hamiltonian, $H_Q$,
 acting between ground states of the classical
Ising model,
\beq
H_Q = - \Gamma \sum_p (\sigma^+ \sigma^- \sigma^+ \sigma^+ + {\rm h.c.})
\eeq
where $p$ denotes a sum over non-crossed plaquettes of the lattice, and
$\sigma^{+(-)}$\ are the raising(lowering) operators of the Ising spins
$\sigma^z=\pm1$.
To study this Hamiltonian it is convenient to use an imaginary time
discrete representation of the path integral via the standard Trotter-Suzuki
procedure -- 
in this case it yields a set of ferromagnetically stacked
planes of the \slwc, with an implicit time continuum limit at large
ferromagnetic coupling, $K^\tau\sum_{n}\sigma^z_n\sigma^z_{n+1}$, 
between 
neighbouring sites in adjacent 
layers, labeled by $n$.\cite{kagtrfield}

We digress briefly to study the imaginary time representation at 
{\it weak} coupling, $K^\tau\ll1$,
 which could be interpreted as the classical
statistical mechanics of a set of square ice planes with a potential
interaction between the planes. In this limit the interaction competes
with the entropy of the planes and we may ask if it prevails despite 
its weakness. To answer this we switch to the height representation 
description of the ice problem. Herein the coarse grained heights
in a given plane are weighted by the pseudo-Boltzmann 
factor:\cite{abraham}
\beq
\rho[h({\bf x})] \propto e^{-\frac{\pi}{12} \int d^2x (\nabla h)^2}
\label{htweight}
\eeq
while the spins (arrows) are represented by $\nabla \times h$ for
their small momentum components and by $e^{i \pi h}$ for momenta
in the vicinity of $(\pi,\pi)$. For uncoupled planes, the distribution
$\prod_n \rho[h_n({\bf x})]$ is a fixed point of a two-dimensional
renormalization group (RG) transformation in the standard fashion. The
coupling between the planes induces two perturbations in the height 
language: the first of these,
$\phi_1=\sum_n \int d^2x (\nabla h_n)\cdot
 (\nabla h_{n+1})$, is exactly marginal
under the RG while the second 
$\phi_2=\sum_n \int d^2x 
\cos [\pi (h_{n+1} - h_n)]$ is irrelevant as is readily
verified. It follows then that the system exhibits a sliding phase
at weak coupling, whence sliding ice.

Returning to the strong coupling problem, it is clear that we need to
look for a phase transition out of the sliding phase into a flat phase
where the layers lock. To this end we have carried out a Monte Carlo
simulation whose results confirm the existence of the sliding phase as
well as of the flat phase
(Fig.~\ref{fig:SxofT}). From our RG analysis, 
this should happen via a second order transition driven by $\phi_2$\ 
becoming relevant, although we have not explored this in detail.
Translated back the the quantum ice problem our
results indicate that the system exhibits long range order of the
antiferroelectric (or F model) kind (Fig.~\ref{fig:sqcrossofS}), 
in which it maximizes the density
of flippable plaquettes.

For the spins on the \slwc\ lattice this is N\'eel order. The quantum
dynamics is the leading order effect of either a transverse field
(in which context the N\'eel phase was conjectured by us 
previously)\cite{kagtrfield} or
an XY exchange of either sign. One may wonder whether this conflicts
with our previous analysis in suggesting N\'eel order at the Heisenberg
point. However it has been shown within spin wave theory\cite{canalswave}
that the N\'eel 
state on the \slwc\ is unstable for $S\leq1$\ so there
is every reason to expect 
a phase transition {\it en route}. For such order by disorder 
phenomena this is by no means exceptional; one generically finds that 
fluctuations generate an ordered state out of a disordered ensemble, 
which they then in turn destroy as their strength increases. 

In summary, we have explored quantum frustrated antiferromagnetism on
the \slwc. Despite an enormous classical ($S=\infty$) degeneracy, we
find a robust ordering tendency for Heisenberg magnets of any spin;
for stacked square ice (Ising spins), a sliding phase precedes an
antiferroelectric ordering transition.

\unue{Acknowledgements:} We are grateful to J.-B. Fouet, E. Lieb,
S. Palmer, S. Sachdev, S. Shastry and P. Sindzingre for useful
discussions, and to C. Henley and C. Lhuillier also for comments on
the manuscript.  We thank P. Chandra for collaboration on related work
and P. Schupp for the pyrochlore figure. This work was supported in
part by the NSF grant No. DMR-9978074, the A.\ P.\ Sloan Foundation
and the David and Lucille Packard Foundation.


\begin{references}
\bibitem{fn-complete} One also has to worry about whether all
the dimer states are linearly independent as the derivation of the QDM
involves inverting their overlap matrix. However, our final result is
unaltered if one includes the diagonal bonds in the derivation of the
dimer Hamiltonian without allowing dimers to be placed on them, for
which case linear independence is assured.
\end{references}
\end{document}